\documentclass[twocolumn,showpacs,preprintnumbers,amsmath,amssymb]{revtex4}
\usepackage{bm}

\newcommand{\ket}[1]{\displaystyle{|#1\rangle}}
\newcommand{\bra}[1]{\displaystyle{\langle #1|}}
\newcommand{\ob}{\mathcal{O}}
\newcommand{\mes}{\mathcal{M}^j}
\newcommand{\pr}{\mathcal{P}}
\newcommand{\wt}[1]{\widetilde{#1}}
\newcommand{\ve}[1]{\mathbf{#1}}
\DeclareMathOperator{\Tr}{Tr}
\newtheorem{theorem}{Theorem}
\newtheorem{lemma}{Lemma}

\begin{document}
\preprint{preprint}
\title{Building an entanglement measure on physical ground}
\author{D.~Teresi, A.~Napoli, A.~Messina}
\affiliation{CNISM, MIUR and Dipartimento di Scienze Fisiche ed
Astronomiche,
via Archirafi 36, 90123 Palermo, Italy, \\
E-mail: messina@fisica.unipa.it }
%\date{\today}
\begin{abstract}
We introduce on physical grounds a new measure of multipartite
entanglement for pure states. The function we define is
discriminant and monotone under LOCC and moreover can be 
expressed in terms of observables of the system.
\end{abstract}

%\pacs{pacs}

\maketitle

Since the beginning of quantum mechanics, entanglement revealed to
be a key concept for the understanding of the nature. Its link
with the foundations of physics was immediately recognized, in
particular in connection with the nonlocality property of quantum
theory. In the last years moreover the interest toward this
fundamental concept of Quantum Mechanics has grown also  in view
of its central role in many fields of contemporary physics, like
quantum information theory or condensed matter physics. In the
last decade, in particular, the fundamental question concerning
how to \emph{quantify} entanglement has received a lot of
attention. To this end different measures of entanglement have
been proposed with respect, in particular, to bipartite systems.
On the contrary entanglement in multipartite systems remains an
open and debated problem. In view of the complexity of such
systems it cannot indeed be understood simply extending the tools
adopted when bipartite entangled states are studied.

Consider a multipartite system composed by $N$ not necessarily
identical subsystems each one living in a finite dimensional
Hilbert space. In this letter we introduce a new measure of
entanglement for such systems in pure states called
\textit{General Entanglement} (GE). This quantity provides a
measure of the entanglement present in the system independently on
how it is distributed among the finitely many possible subsystems.
The GE proves to be easily computable and reduces to Meyer and
Wallach's Global Entanglement \cite{Meyer} when qubit systems are
considered. A very important aspect is that the quantity we
introduce has an immediate interpretation. Making indeed physical
considerations of clear meaning we construct our new measure
function directly starting from the concept of separability. The
quantity we define is moreover characterized by many appealing
properties making it very attractive both from a conceptual and an
experimental point of view. As well known a pure state of a
multipartite system is said to be completely separable if it can
be written as tensor product of states of each subsystem. At the
same time a state is separable with respect to an assigned
subsystem if, and only if, no physical quantity of the subsystem
under scrutiny can be changed acting on the rest of the system.
Let thus consider a multipartite system in a pure state
$\ket{\psi}$ and focus on the single $j$-th subsystem. A
projective measurement \cite{note1,LeBellac} on the rest of the
system is defined as:
\begin{equation}\label{eq:mes}
\mes =
\bigl\{\pr_i=\ket{\chi_i}\bra{\chi_i}\bigr\}
\end{equation}
with
\begin{equation}\label{eq:P_operators}
\sum_i \pr_i = \mathbb{I} \;\;\;\;\text{and}\;\;\;\; \pr_i =
\mathbb{I}^{(j)}\otimes \pr_i^{(r)}
\end{equation}
In eq. (\ref{eq:P_operators}) the projection operators $\pr_i$ act
on the Hilbert space of the total system whereas $\pr_i^{(r)}$ act
on the Hilbert space relative to the system obtained excluding the
$j-$th subsystem from the total one.

As a result of the measurement the system initially in the state
$\rho\equiv\ket{\psi}\bra{\psi}$ is projected, with probability
$p_i$, onto the pure state $\rho_i$ corresponding to the obtained
outcome: $\rho\xrightarrow{\phantom{i}\mes\phantom{i}}\{p_i,
\rho_i\}$. Thus, whatever the observable $\ob = \ob^{(j)} \otimes
\mathbb{I}^{(r)}$ is, the quantity:
\begin{equation}\label{eq:R_def}
R_{\ob}^{(j), \pr_i}\bigl(\rho\bigr) = \bigl( \Tr{\rho\ob} -
\Tr{\rho_i\ob} \bigr)^2
\end{equation}
is zero if $\rho$ is separable with respect to the subsystem $j$.
This statement is in addition true whatever  the chosen projective
measurement $\mes$ is. If, on the contrary, the quantity
$R_{\ob}^{(j), \pr_i}(\rho)$ is zero for any $\ob^{(j)}$, $\mes$
and for any outcome $i$, we may claim with certainty that the
state of the $j$-th subsystem is not correlated with the rest of
the system in any way. Under this condition $\rho$ must be
separable with respect to the subsystem $j$; thus, if this
property is true for every subsystem, the state must be completely
separable. Guided from these considerations we introduce the
quantities:
\begin{equation}\label{eq:Emj}
\mathcal{E}_{\mes}^{(j)}\bigl(\rho\bigr) =
\sum_i\,p_i\,\max_{\ob\in\Omega} R_{\ob}^{(j), \pr_i}(\rho)
\end{equation}
where $\Omega$ is the set of all the observables $\ob^{(j)}
\otimes \mathbb{I}^{(r)}$, $\ob^{(j)}$ acting on the state space
of the $j$-th subsystem. By definition this quantity gives an
estimation of the \textit{average departure} from the separability
condition.
 $\mathcal{E}_{\mes}^{(j)}(\rho)$ is indeed
equal to zero with certainty only if the state is separable with
respect to $j$. Let's however observe that it goes to infinity in
the opposite case being $R_{\alpha\ob}^{(j), \pr_i}(\rho)=\alpha^2
R_{\ob}^{(j), \pr_i}(\rho), \; \alpha\in\mathbb{R}$. Confining
ourselves however to the set $\Omega$ of all the normalized
observables, with respect to a prefixed norm, the quantity
$\mathcal{E}_{\mes}^{(j)}(\rho)$ becomes finite. Let's moreover
observe that since $R_{\ob}^{(j),
\pr_i}(\rho)=R_{\tilde{\ob}}^{(j), \pr_i}(\rho)$ with
$\tilde{\ob}=\ob - r\mathbb{I}, r\in\mathbb{R}$, $R_{\ob}^{(j),
\pr_i}(\rho)$ does not depend on $\Tr{\ob}$. Thus, without loss of
generality we put
\begin{equation} \label{eq:omega_norm}
\Omega = \bigl\{\ob = \ob^{(j)} \otimes \mathbb{I}^{(r)},
\ob^\dag=\ob, \Tr{\ob}=0, \|\ob\|=1\bigr\}
\end{equation}

The quantity $\mathcal{E}_{\mes}^{(j)}(\rho)$ evaluated in the set
$\Omega$ defined by \eqref{eq:omega_norm}, gives an estimation of
the degree of entanglement existing between the $j$-th subsystem
and and the rest of the system. In other words the greater
$\mathcal{E}_{\mes}^{(j)}(\rho)$ is, the greater is the influence
on the $j$-th subsystem stemming from the measurement on the rest
of the system. Thus, when a system composed by $N$ subsystems is
in a pure state $\rho\equiv\ket \psi\bra\psi$, we are naturally
guided to adopt as a measurement of entanglement the quantity
\begin{equation}\label{eq:def}
E_g\bigl(\rho\bigr) = \frac{1}{N} \sum_{j=1}^N \max_{\mes}
\mathcal{E}_{\mes}^{(j)}(\rho)
\end{equation}
In what follows we will refer to this measure as \emph{General
Entanglement}. By definition $E_g\bigl(\rho\bigr)$ may be
evaluated for pure states of arbitrarily large multipartite
systems, whose $N$ constituents have finite dimensional Hilbert
spaces. It is important to stress that, differently from the
Global entanglement of Meyer and Wallach \cite{Meyer} or its
generalizations proposed by Rigolin et al. \cite{Rigolin}, our
definition does not require that such Hilbert spaces have the same
dimensions. We now prove that GE is a \emph{good} entanglement
measure for pure states \cite{Vedral, Vidal, Horodecki, Donald}.
To this end we begin demonstrating the following
\begin{theorem}
General Entanglement is discriminant, that is $E_g(\rho)=0
\Longleftrightarrow \rho$ is completely separable.
\end{theorem}
\emph{Proof:} $\rho$ completely separable implies $E_g(\rho)=0$
being $R_{\ob}^{(j), \pr_i}(\rho)=0$ whatever the observable $\ob$
and the subsystem $j$ are. Conversely, $E_g(\rho)=0$ implies
$\mathcal{E}_{\mes}^{(j)}(\rho)=0$ whatever $j$ and $\mes$ are.
Since in addition in correspondence to an outcome $i$ with
$p_{i}\neq 0$ $\Tr{\rho\ob} = \Tr{\rho_i\ob}$ for any $\ob$, then
$\rho^{(j)}\equiv Tr_r{\rho}$ and $\rho_i^{(j)}\equiv
Tr_r{\rho_i}$ coincide. But, $\rho_i$ is pure and separable with
respect to $j$; thus $\rho_i^{(j)}$, and therefore $\rho^{(j)}$,
is pure too. Then $\rho$ is separable with respect to $j$. Since
such a property holds for any subsystem $j$, the state
$\rho\equiv\ket \psi\bra\psi$ is completely separable. $\Box$

Another remarkable features of our GE is its invariance under
local unitary operations. It is indeed possible to prove the
following
\begin{theorem}
$E_g(\rho) = E_g(U\rho U^\dag)$, with $U^\dag = U^{-1}$ and
\mbox{$U = U^{(1)} \otimes U^{(2)} \otimes \dots \otimes U^{(N)}$}
\end{theorem}
\emph{Proof:} Putting $\wt{\rho} = U \rho U^\dag$ for every
admissible measurement $\mes=\{\pr_i\}$ consider the
\emph{transformed} measurement $\wt{\mes}\equiv\{\wt{\pr_i} = U
\pr_i U^\dag \}$ satisfying \eqref{eq:mes}).  It is immediate to
convince oneself that since
\begin{equation}
\wt{p_i} = \Tr{\wt{\rho} \wt{\pr_i}} = p_i \qquad \wt{\rho_i} =
 \frac{1}{\wt{p_i}} \wt{\pr_i} \wt{\rho} \wt{\pr_i} = U \rho_i U^\dag
\end{equation}
then
\begin{equation}
R_{U \ob U^\dag}^{(j), \wt{\pr_i}}(\wt{\rho}) = R_{\ob}^{(j), \pr_i}(\rho)
\end{equation}
so that
\begin{equation} \label{eq:maxR}
\max_{\ob\in\Omega} R_{\ob}^{(j), \wt{\pr_i}}(\wt{\rho}) =
\max_{\ob\in\Omega} R_{\ob}^{(j), \pr_i}(\rho)
\end{equation}
Thus $\mathcal{E}_{\wt{\mes}}^{(j)}(\wt{\rho}) =
\mathcal{E}_{\mes}^{(j)}(\rho)$ and, therefore, $E_g(\rho) =
E_g(U\rho U^\dag)$. $\Box$

To obtain an explicit expression for the GE we normalize the
observables of the set $\Omega$ with respect to the \emph{trace
scalar product}:
\begin{equation}\label{eq:scal_pro}
( A, B ) = \Tr( A B ) \Longrightarrow \|A\|^2 = \Tr A^2
\end{equation}
This choice of the set $\Omega$ allows us to prove that GE is
monotone under LOCC. To demonstrate this remarkable property it is
convenient to prove in advance the following general Lemma:
\begin{lemma}\label{lem:bas}
Let $\{A_k\}$ be an orthonormal basis  (with respect to
\eqref{eq:scal_pro}) in the vectorial space of traceless hermitian
$D \times D$ matrices. For every $D \times D$ hermitian matrix
$\sigma$ with $Tr\sigma=1$ we have:
\begin{equation}
\sum_k (\Tr \sigma A_k)^2 \equiv\sum_k \langle A_k
\rangle_\sigma^2 =\Tr \sigma^2 - \frac{1}{D}
\end{equation}
\end{lemma}
\emph{Proof:} Expanding $\sigma$ in the basis $\{\mathbb{I},
A_k\}$
\begin{equation}\label{eq:decomp}
\sigma = \frac{1}{D} \mathbb{I} + \sum_k r_k A_k
\end{equation}
we obtain:
\begin{equation}\sum_k (\Tr \sigma A_k)^2 = \sum_k r_k^2 \qquad \Tr \sigma^2 =
\frac{1}{D} + \sum_k r_k^2 \quad
\end{equation}$\Box$

Let us now focus on a single subsystem $j$ and indicate by
$D^{(j)}$ the dimension of its Hilbert space. Consider an
orthonormal set of $(D^{(j)})^2-1$ traceless observables $\{A_k\}$
relative to the $j$-th subsystem. Whatever the observable $\ob
\equiv \ob^{(j)}\otimes \mathbb{I}^r \in \Omega$ is we can write
$\ob^{(j)} = \sum_k o_k A_k$, with $\sum_k o_k^2 = 1$. For
simplicity, in what follows we write $\ob^{(j)} = \ve{\hat{o}}
\cdot \ve{A}$ with $\ve{\hat{o}}=(o_1, o_2, ...)$ and
$\ve{A}=(A_1, A_2,...)$, and denote by
$\langle\ve{A}\rangle_{\rho}$ the vector of components $\langle
A_k \rangle_{\rho}\equiv Tr{(\rho A_k\otimes\mathbb{I}^r)}$ in
$\mathbb{R}^{(D^{(j)})^2-1}$. Exploiting this notation, eq.
\eqref{eq:R_def} may be cast in the form
\begin{equation}
R_{\ob}^{(j), \pr_i}(\rho) = \bigl[\ve{\hat{o}} \cdot \bigl(
\langle\ve{A}\rangle_{\rho_i} - \langle\ve{A}\rangle_{\rho} \bigr)
\bigr]^2
\end{equation}
Observing that the set $\Omega$ can be obtained simply varying the
unit vector $\ve{\hat{o}}$ we may write
\begin{equation}
\mathcal{E}_{\mes}^{(j)}(\rho) = \sum_i\,p_i\, \bigl(
\langle\ve{A}\rangle_{\rho_i} - \langle\ve{A}\rangle_{\rho} \bigr)
\cdot \bigl( \langle\ve{A}\rangle_{\rho_i} -
\langle\ve{A}\rangle_{\rho} \bigr)
\end{equation}

Taking into consideration the fact that $\rho_i^{(j)}$ is pure and
using Lemma~\ref{lem:bas}, we have:
\begin{equation}
\mathcal{E}_{\mes}^{(j)}(\rho) = 1 - \frac{1}{D^{(j)}} +
\langle\ve{A}\rangle_{\rho} \cdot \langle\ve{A}\rangle_{\rho} - 2
\langle\ve{A}\rangle_{\rho} \cdot
\sum_i\,p_i\,\langle\ve{A}\rangle_{\rho_i}
\end{equation}
Starting from eq. \eqref{eq:mes} it's easy to prove that:
\begin{equation}
\sum_i\,p_i\,\langle\ve{A}\rangle_{\rho_i} = \langle\ve{A}\rangle_{\rho}
\end {equation}
and thus
\begin{equation}\label{eq:eps__A}
\mathcal{E}_{\mes}^{(j)}(\rho) = 1 - \frac{1}{D^{(j)}} -
\langle\ve{A}\rangle_{\rho} \cdot \langle\ve{A}\rangle_{\rho}
\end{equation}

Summing up, with the choice \eqref{eq:scal_pro}, the quantities
$\mathcal{E}_{\mes}^{(j)}(\rho)$ do not depend on the measures
$\mes$ and the maximization in \eqref{eq:def} becomes trivial. We
wish moreover point out that in view of Lemma~\ref{lem:bas}, $0
\leq \sum_k \langle A_k \rangle_\rho^2 \leq 1-\frac{1}{D^{(j)}}$.
This inequality suggests to rescale
$\mathcal{E}_{\mes}^{(j)}(\rho)$ as follows
$\mathcal{E}_{\mes}^{(j)}(\rho)\longrightarrow
(1-\frac{1}{D^{(j)}})^{-1}\mathcal{E}_{\mes}^{(j)}(\rho)$
obtaining $\mathcal{E}_{\mes}^{(j)}(\rho) = 1 -
\frac{D^{(j)}}{D^{(j)}-1}\langle\ve{A}\rangle_{\rho} \cdot
\langle\ve{A}\rangle_{\rho}$. Thus we may define the normalized
General Entanglement as:
\begin{align}\label{eq:gen_ent}
&E_g(\rho) = 1 - \frac{1}{N} \sum_j \frac{D^{(j)}}{D^{(j)}-1} \,
\langle\ve{A}^{(j)}\rangle_{\rho} \cdot \langle\ve{A}^{(j)}\rangle_{\rho} =\notag\\
&= 1 + \frac{1}{N} \sum_j \frac{1}{D^{(j)}-1} - \frac{1}{N} \sum_j
\frac{D^{(j)}}{D^{(j)} - 1} \Tr (\rho^{(j)})^2
\end{align}
Thus with the choice \eqref{eq:scal_pro} GE reveals to be related
to the average purity of the state, and, when we deal with equal
dimensional subsystems it reduces to the generalized global
entanglement $E_g^{(1)}$ \cite{Rigolin}. Moreover under the choice
\eqref{eq:scal_pro} the GE is surely monotone as proved by the
following theorem:
\begin{theorem}\label{the:mono}
General Entanglement is monotone under LOCC if we make the choice
\eqref{eq:scal_pro}.
\end{theorem}
\emph{Proof:} In order to prove this statement it is sufficient to
demonstrate that $-\sum_j \frac{D^{(j)}}{D^{(j)} - 1} \Tr
(\rho^{(j)})^2$ does not increase under LOCC. Let us focus on the
subsystem $j$, and consider the bipartition (\emph{$j$ - rest of
the system}). With respect to this bipartition the state admits
Schmidt decomposition and, using Schur's theorem
\cite{geometry_Schur}, it is easy to prove that $\Tr
(\rho^{(j)})^2$ is a Schur-convex function of the Schmidt
coefficients; then, thanks to Nielsen's majorization theorem
\cite{Nielsen}, $-\Tr (\rho^{(j)})^2$ does not increase under LOCC
with respect to the bipartition. But a LOCC with respect to
\emph{all} partitions is a LOCC with respect to the fixed
bipartition; so $-\sum_j \frac{D^{(j)}}{D^{(j)} - 1} \Tr
(\rho^{(j)})^2$ is not-increasing under LOCC. $\Box$

The ability of writing the General Entanglement $E_g(\rho)$ as
expressed by eq. (\ref{eq:gen_ent}) is remarkable not only because
it allows us to prove its monotonicity but also in view of the
following considerations. First of all eq. (\ref{eq:gen_ent})
clearly shows that the GE is a linear function of the "average"
purity of the state of all the subsystems. Moreover, exploiting
the first equality of eq. (\ref{eq:gen_ent}), we may express
$E_g(\rho)$ in terms of mean values of local observables. This
circumstance is of particular relevance from an experimental point
of view giving the possibility of testing directly in laboratory
the quantity $E_g(\rho)$ here defined. In what follows we will
apply the new concept of GE in order to evaluate the degree of
entanglement of assigned multipartite systems. Let's begin
considering a system of $N$ spin $\frac{1}{2}$. In correspondence
to each subsystem the operators $S_x$, $S_y$, and $S_z$ are
traceless and orthogonal each other so that, once normalized they
provide the following useful set of operators:
\begin{align}
A_1 &\equiv \sqrt{2} S_z = \frac{1}{\sqrt{2}}\begin{pmatrix} 1 & 0 \\ 0 & -1\end{pmatrix}  \notag\\
A_2 &\equiv \sqrt{2} S_x = \frac{1}{\sqrt{2}} \begin{pmatrix} 0 & 1 \\ 1 & 0 \end{pmatrix}\\
A_3 &\equiv \sqrt{2} S_y = \frac{1}{\sqrt{2}} \begin{pmatrix} 0 &
-\imath \\ \imath & 0  \notag \end{pmatrix}
\end{align}
with $\hbar = 1$. Exploiting \eqref{eq:gen_ent}, it is immediate
to conclude that in the case under scrutiny the degree of
multipartite entanglement measured by GE is simply given by
\begin{equation}
E_g\bigl(\rho\bigr) = 1 - \frac{4}{N} \sum_j
\,\langle\mspace{1.0mu}\ve{S}^{(j)}\mspace{1.0mu}\rangle^2
\end{equation}
This expression coincides with the Meyer-Wallach global
entanglement \cite{Meyer,Brennen,Amico} when $N$ qubits are
considered.  It is of relevance to observe that if $N=2$ GE can be
directly related to the concurrence function $C$ \cite{Wootters}
being in particular $E_g\bigl(\rho\bigr) =
C^2\bigl(\ket{\psi}\bigr)$.

Suppose now that the system of interest is composed by $N$ spins
$1$. In this case in order to construct the appropriate set of
$\{A_k^{(j)}\}$ operators let's start by considering the following
linearly independent observables:
\begin{align}
S_z &= \begin{pmatrix}  1 & 0 & 0 \\
                        0 & 0 & 0 \\
                        0 & 0 & -1 \end{pmatrix} &
S_x &= \frac{1}{\sqrt{2}} \begin{pmatrix}   0 & 1 & 0 \\
                                            1 & 0 & 1 \\
                                            0 & 1 & 0 \end{pmatrix} \notag\\
S_y &= \frac{1}{\sqrt{2}} \begin{pmatrix}   0 & -\imath & 0 \\
                                            \imath & 0 & -\imath \\
                                            0 & \imath & 0 \end{pmatrix} &
S_{xy} &= S_x S_y + S_y S_x \notag\\
S_{xz} &= S_x S_z + S_z S_x & S_{yz} &= S_y S_z + S_z S_y \notag\\
S_x^2 &= S_x S_x & S_y^2 &= S_y S_y
\end{align}
Orthonormalizing this set by the Gram-Schmidt method
\cite{Arfken}, we obtain the following orthonormal traceless
basis:
\begin{align}
A_1 &= \frac{1}{\sqrt{\,2}} \, S_z & A_2 &= \frac{1}{\sqrt{\,2}} \, S_x \notag\\
A_3 &= \frac{1}{\sqrt{\,2}} \, S_y & A_4 &= \frac{1}{\sqrt{\,2}} \, S_{xy} \notag\\
A_5 &= \frac{1}{\sqrt{\,2}} \, S_{xz} & A_6 &= \frac{1}{\sqrt{\,2}} \, S_{yz} \notag\\
A_7 &= \sqrt{\frac{3}{2}}\,S_x^2 - \sqrt{\frac{2}{3}}\,\mathbb{I} &
A_8 &= \sqrt{2} \, S_y^2 + \frac{1}{\sqrt{\,2}} \, S_x^2 - \sqrt{2} \, \mathbb{I}
\end{align}
Thus \eqref{eq:gen_ent} becomes:
\begin{equation}
E_g\bigl(\rho\bigr) = 1 - \frac{3}{2 N} \sum_j
\,\langle\mspace{1.0mu}\ve{A}^{(j)}\mspace{1.0mu}\rangle^2_\rho
\end{equation}

As we have previously stressed, GE does not distinguish between
``truly'' $N$-partite entanglement \cite{Horodecki_review} and
partially separable entanglement. In other words the quantity
$E_g(\rho)$ is different from zero also in correspondence to a
state separable with respect to some bipartition. It indeed
indicates how much \emph{global} entanglement is present in the
system. Anyway, if we are interested only in $N$-partite
entanglement, a variant of GE can be introduced. Let us denote by
$\rho_P$ the state of the system thought as a bipartite system
induced by the bipartition $P$. Then, the following measure:
\begin{equation}\label{eq:Eg_N}
E_g^N\bigl(\rho\bigr) = \min \Bigl\{ E_g\bigl(\rho\bigr),
E_g\bigl(\rho_P\bigr), \forall\;\text{bipartition}\; P\, \Bigr\}
\end{equation}
is nonzero if, and only if, the state is $N$-partite truly
entangled, and is less or equal than $E_g\bigl(\rho\bigr)$. If the
system is not too large, the quantity $E_g^N\bigl(\rho\bigr)$
defined by \eqref{eq:Eg_N} is simple to compute. In addition it is
monotone in view of the fact that a LOCC with respect to all the
subsystems is a LOCC with respect to a bipartition.

Summarizing, in this paper we propose a new way to quantify
entanglement in multipartite pure systems. In contrast to the
Global Entanglement \cite{Meyer} and its generalizations
\cite{Rigolin}, our measure does not require that all the
subsystems have the same dimension. Thus GE can be applied to more
general physical situations. In addition the measure we propose
reveals to be a good one being discriminant, invariant under local
unitary operations and monotone under LOCC at least when the
normalization (10) is adopted. Moreover the possibility of
expressing GE in terms of mean values of suitable local quantities
turns out to be very attractive from the experimental point of
view. By definition the quantity we introduce does not allow to
distinguish the many ways in which a multipartite system can be
entangled. Our aim is indeed to quantify the entanglement present
in a multipartite system independently of its distribution. On the
other hand the generalization of GE proposed in eq. (25) allows us
to distinguish genuine multipartite entangled states. The
fundamental aspect of our GE is the fact that it is constructed
following a quite simple reasoning based on physical grounds. This
directly furnish the possibility of interpreting our function in a
clear way. The starting point is that more our physical
predictions on a subsystem are changeable acting on the rest of
the system, more the subsystem is entangled with the rest. An
important result, from the conceptual point of view, is that GE
improves, with respect to the notion of monotonicity under LOCC,
our capability to \emph{physically} say that a state is more or
less entangled than another. In fact, as far as monotonicity, if a
state $\ket{\psi}$ can be transformed into $\ket{\phi}$ by LOCC,
we physically say that $\ket{\psi}$ is more (or equal) entangled
than $\ket{\phi}$. But the order imposed by LOCC is only partial;
so, let us consider two states that cannot be converted into each
other. We could not physically say that a state is more entangled
than the other, if we limit the concept of entanglement to a
quantity that does not increase under LOCC. The physical meaning
of GE provides a way to compare, on physical basis, the
entanglement of such states \cite{note2}. Thanks to the fact that
GE is monotone, this physical meaning is not in contrast with the
commonly accepted fact that entanglement is a quantity that does
not increase under LOCC.

Concluding we wish to stress that at least in principle the
definition of GE could be extended to the case of
infinite-dimensional subsystems, provided that all the involved
summations converge. The most delicate point is that the
normalization \eqref{eq:scal_pro} does not work in this case. As
far as a possible generalization to statistical mixtures, let us
observe that in these cases in order to estimate the degree of
entanglement we can take the convex-roof \cite{Uhlmann} of GE. In
other words we can adopt the following quantity $E(\rho) = \inf
\sum_i p_i\,E_g\bigl(\ket{\psi_i}\bigr)$, where the infimum is
taken over all the possible decompositions $\rho = \sum_i p_i\,
\ket{\psi_i}\bra{\psi_i}$. However, this quantity is not easily
computable, because of the involved extremization.


\begin{thebibliography}{10}
\bibitem{note2} From the practical point of view this
comparison is made possible by the fact that General Entanglement
is simple to compute.
\end{thebibliography}
 \end{document}